\pdfoutput=1
\documentclass[conference]{IEEEtran}
\IEEEoverridecommandlockouts
\usepackage{cite}
\usepackage{amsmath,amssymb,amsfonts}
\usepackage{graphicx}
\usepackage{float}
\usepackage{dblfloatfix}
\usepackage{textcomp}
\usepackage[dvipsnames]{xcolor}
\usepackage{siunitx}
\usepackage{xfrac}
\usepackage{booktabs} 
\usepackage{multirow} 
\usepackage{subcaption}
\usepackage{stackengine} 

\def\BibTeX{{\rm B\kern-.05em{\sc i\kern-.025em b}\kern-.08em
T\kern-.1667em\lower.7ex\hbox{E}\kern-.125emX}}

\usepackage{scalerel}
\usepackage{tikz}
\usetikzlibrary{svg.path}

\definecolor{orcidlogocol}{HTML}{A6CE39}
\tikzset{
	orcidlogo/.pic={
		\fill[orcidlogocol] svg{M256,128c0,70.7-57.3,128-128,128C57.3,256,0,198.7,0,128C0,57.3,57.3,0,128,0C198.7,0,256,57.3,256,128z};
		\fill[white] svg{M86.3,186.2H70.9V79.1h15.4v48.4V186.2z}
		svg{M108.9,79.1h41.6c39.6,0,57,28.3,57,53.6c0,27.5-21.5,53.6-56.8,53.6h-41.8V79.1z M124.3,172.4h24.5c34.9,0,42.9-26.5,42.9-39.7c0-21.5-13.7-39.7-43.7-39.7h-23.7V172.4z}
		svg{M88.7,56.8c0,5.5-4.5,10.1-10.1,10.1c-5.6,0-10.1-4.6-10.1-10.1c0-5.6,4.5-10.1,10.1-10.1C84.2,46.7,88.7,51.3,88.7,56.8z};
	}
}

\newcommand\orcidicon[1]{\href{https://orcid.org/#1}{\mbox{\scalerel*{
				\begin{tikzpicture}[yscale=-1,transform shape]
				\pic{orcidlogo};
				\end{tikzpicture}
			}{|}}}}

\usepackage{hyperref} 
\hypersetup{hidelinks}

\begin{document}

\title{
Learning Super-Resolution Ultrasound Localization Microscopy from Radio-Frequency Data
\thanks{
This work is funded by the Hasler Foundation under project number 22027. \\
$^{\star}$Corresponding author email: \href{mailto:christopher.hahne@unibe.ch}{\textcolor{blue}{christopher.hahne [ät] unibe.ch}}
}
}

\author{
\IEEEauthorblockN{Christopher Hahne\textsuperscript{$\star$}
}
\IEEEauthorblockA{
\textit{ARTORG Center} \\
\textit{University of Bern} \\
Bern, Switzerland \\
}

\and
\IEEEauthorblockN{Georges Chabouh
}
\IEEEauthorblockA{
\textit{INSERM}, \textit{CNRS} \\
\textit{Sorbonne Université} \\ 
Paris, France \\
}
\and
\IEEEauthorblockN{
Olivier Couture
}
\IEEEauthorblockA{
\textit{INSERM}, \textit{CNRS} \\
\textit{Sorbonne Université} \\ 
Paris, France \\
}

\and
\IEEEauthorblockN{Raphael Sznitman
}
\IEEEauthorblockA{
\textit{ARTORG Center} \\
\textit{University of Bern} \\
Bern, Switzerland \\
}
}

\maketitle
\begin{abstract}
Ultrasound Localization Microscopy (ULM) enables imaging of vascular structures in the micrometer range by accumulating contrast agent particle locations over time.
Precise and efficient target localization accuracy remains an active research topic in the ULM field to further push the boundaries of this promising medical imaging technology.
Existing work incorporates Delay-And-Sum (DAS) beamforming into particle localization pipelines, which ultimately determines the ULM image resolution capability.
In this paper we propose to feed unprocessed Radio-Frequency (RF) data into a super-resolution network while bypassing DAS beamforming and its limitations. 
To facilitate this, we demonstrate label projection and inverse point transformation between B-mode and RF coordinate space as required by our approach.
We assess our method against state-of-the-art techniques based on a public dataset featuring in silico and in vivo data.
Results from our RF-trained network suggest that excluding DAS beamforming offers a great potential to optimize on the ULM resolution performance.
\end{abstract}
\section{Introduction}

\begin{figure}[!t]
    \centering
    \includegraphics[width=.9\linewidth]{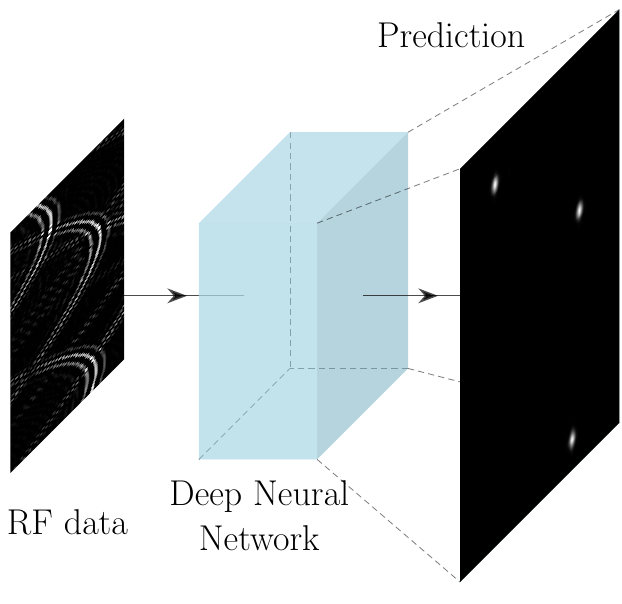}
    \caption{\textbf{RF-ULM pipeline:} Our proposed approach involves utilizing raw Radio-Frequency (RF) data as input to a \mbox{super-resolution} neural network to generate a localization frame without the need for delay and sum beamforming. The localized points are mapped to \mbox{B-mode} coordinate space using an affine transformation.}
    \label{fig:arch}
\end{figure}

In the evolving literature of Ultrasound Localization Microscopy (ULM), a compelling possibility emerges: direct localization without computational beamforming.

ULM has boosted the ultrasound imaging resolution by pinpointing and fusing individual contrast agent particles, 
known as microbubbles, across multiple data frames~\cite{heiles2022pala}. 
Reliable and precise localization of microbubbles thus became the main research topic for ULM in recent years~\cite{heiles2022pala,van2020super,liu2020deep}. So far, it has been widely accepted to use beamformed images as inputs for the localization procedure. However, we scrutinize this common practice by harnessing the data prior to beamforming. 

Our hypothesis builds upon Geometric ULM \mbox{(G-ULM)}~\cite{gulm:2023} where it is assumed that beamforming reduces the signal information, which would deteriorate the localization ability for ULM. This is because Radio-Frequency (RF) data contains rich information such as the reflected wavefront shape, which is irrevocably lost during channel data fusion by the beamforming process.
Notably, ultrasound image formation has recently been accomplished from RF data without beamforming~\cite{nair2018deep,gulm:2023}. For instance, Deep Neural Networks (DNNs) have shown promise to reconstruct objects from raw RF data in the absence of Delay-And-Sum (DAS) beamforming~\cite{nair2018deep}. As an alternative, \mbox{G-ULM} has achieved image recovery through trilateration that maps Time-of-Arrival detections to \mbox{B-mode} coordinate space~\cite{gulm:2023}. These findings motivate us to propose a framework where we feed RF data into a super-resolution DNN as depicted in Fig.~\ref{fig:arch}. %
Our proposed pipeline can be thought of as integrating the adaptive beamforming process into the high-resolution localization network. %
However, leveraging \mbox{RF-trained} networks for ULM requires a coordinate conversion from RF to \mbox{B-mode} space and vice versa. We demonstrate label and point transformation between \mbox{B-mode} and RF coordinate spaces as required by our method. 
We provide a comparison of our proposed approach against state-of-the-art techniques in the field. %
Our presented method achieves competitive localization scores offering a considerable alternative to the traditional beamforming-based ULM.

\section{Method}

\begin{figure*}[b]
  \centering
  \begin{minipage}[b]{0.16\textwidth}
    \centering
    \includegraphics[width=\textwidth,trim=240 538 1030 238, clip]{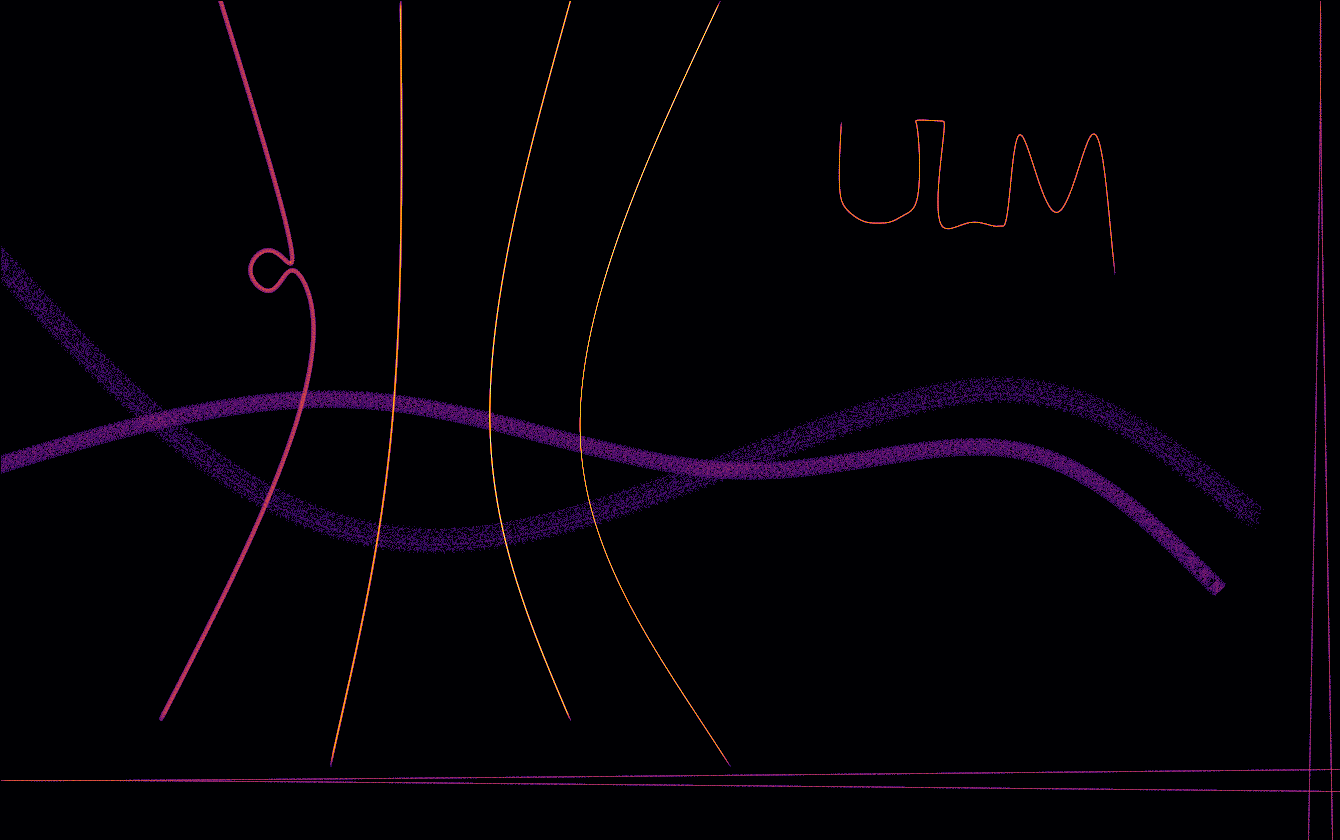}\par\vspace{1mm}
    \includegraphics[width=\textwidth,trim=310 45 980 750, clip]{./img/pala_sim_results/gtru_ulm_img_15k_gam0.9.png}\par\vspace{1mm}
    \includegraphics[width=\textwidth,trim=820 560 200 100, clip]{./img/pala_sim_results/gtru_ulm_img_15k_gam0.9.png}
    \subcaption{Ground Truth\label{fig:method1}}
    \vspace{+.1cm}
  \end{minipage}
  \hfill
  \begin{minipage}[b]{0.16\textwidth}
    \centering
    \includegraphics[width=\textwidth,trim=240 538 1030 238, clip]{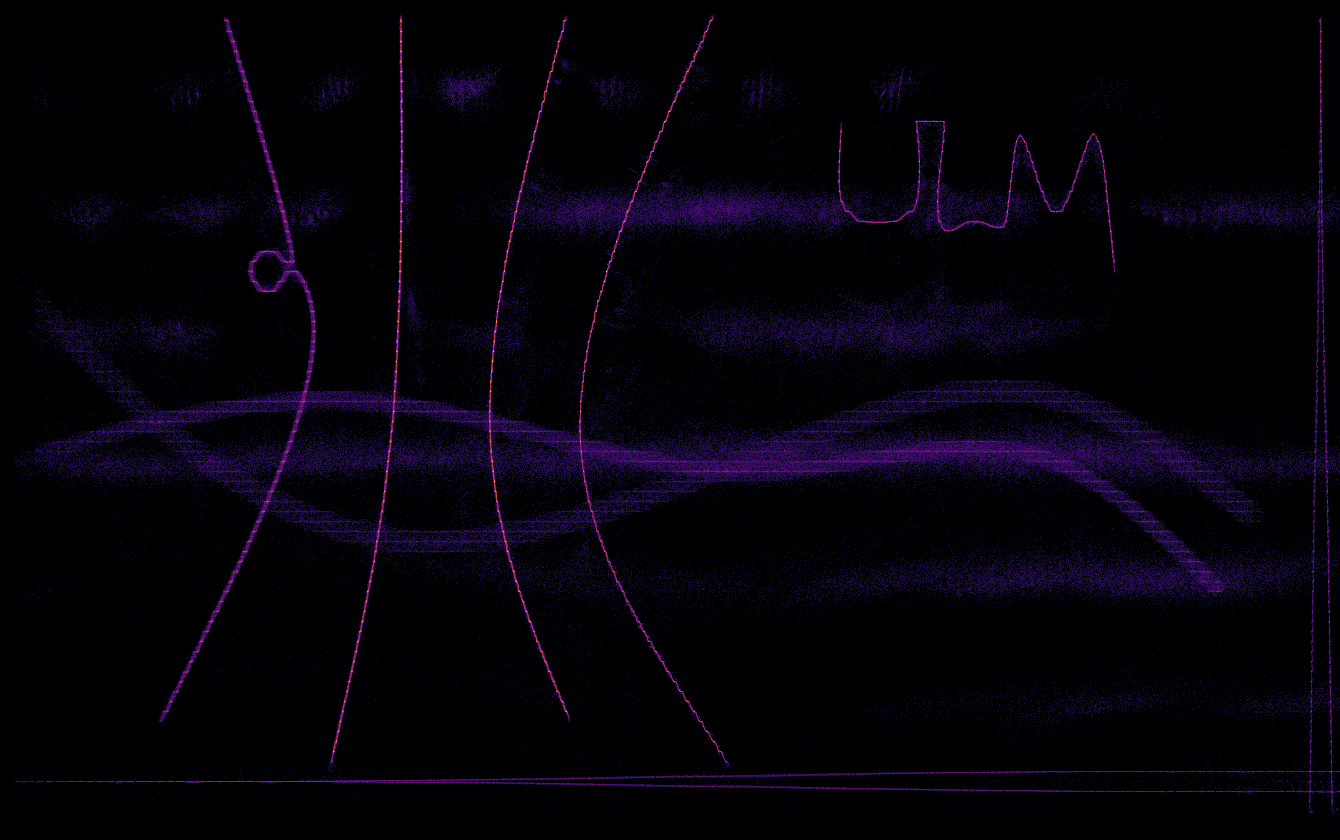}\par\vspace{1mm}
    \includegraphics[width=\textwidth,trim=310 45 980 750, clip]{./img/pala_sim_results/pala_ulm_img_15k_gam0.9_128chs.png}\par\vspace{1mm}
    \includegraphics[width=\textwidth,trim=820 560 200 100, clip]{./img/pala_sim_results/pala_ulm_img_15k_gam0.9_128chs.png}
    \subcaption{RS~\{128\}~\cite{heiles2022pala}\label{fig:method2}}
    \vspace{+.1cm}
  \end{minipage}
  \hfill
	\begin{minipage}[b]{0.16\textwidth}
		\centering
		\includegraphics[width=\textwidth,trim=240 538 1030 238, clip]{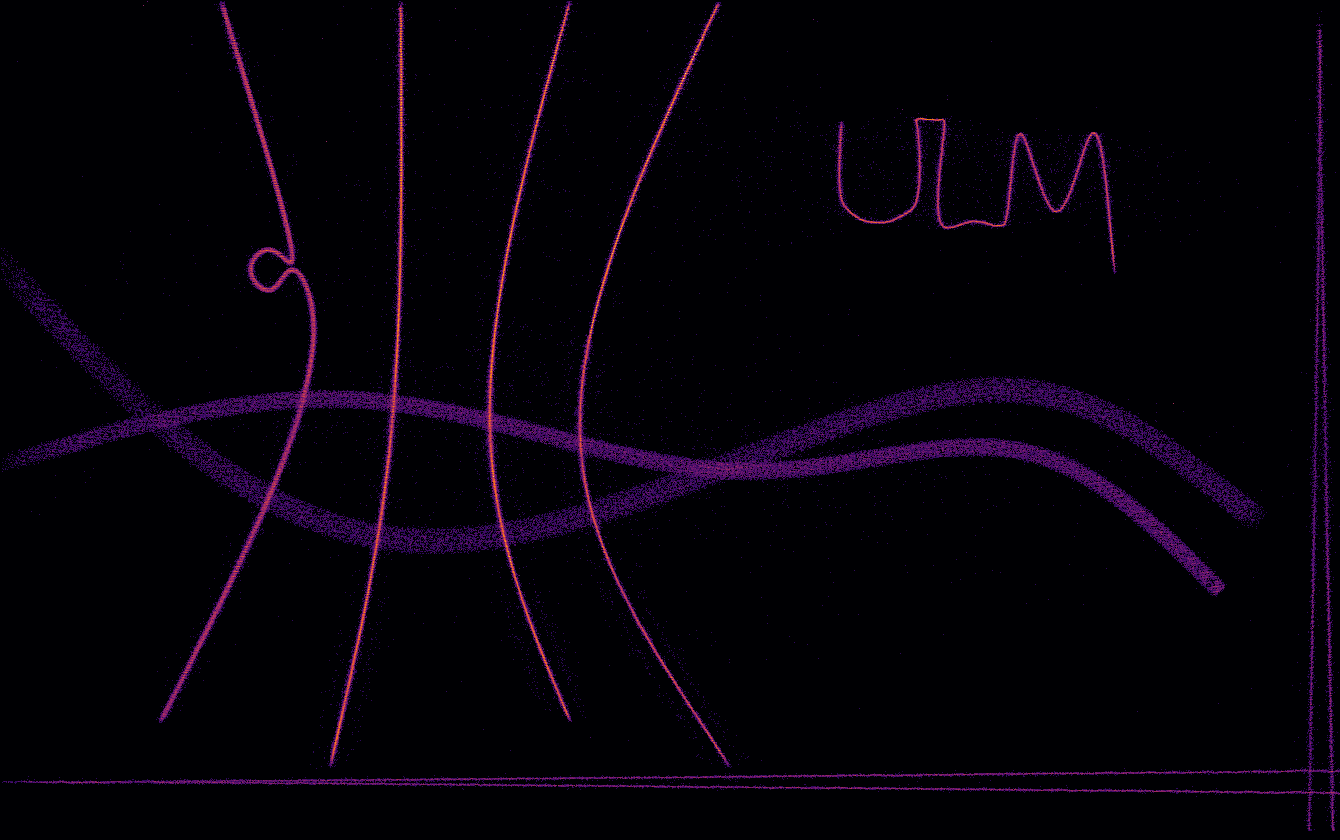}\par\vspace{1mm}
		\includegraphics[width=\textwidth,trim=310 45 980 750, clip]{./img/pala_sim_results/gulm_ulm_img_15k_gam0.9_16chs.png}\par\vspace{1mm}
		\includegraphics[width=\textwidth,trim=820 560 200 100, clip]{./img/pala_sim_results/gulm_ulm_img_15k_gam0.9_16chs.png}
		\subcaption{GULM~\{16\}~\cite{gulm:2023}\label{fig:method3}}
		\vspace{+.1cm}
	\end{minipage}
  \hfill
  \begin{minipage}[b]{0.16\textwidth}
    \centering
    \includegraphics[width=\textwidth,trim=240 538 1030 238, clip]{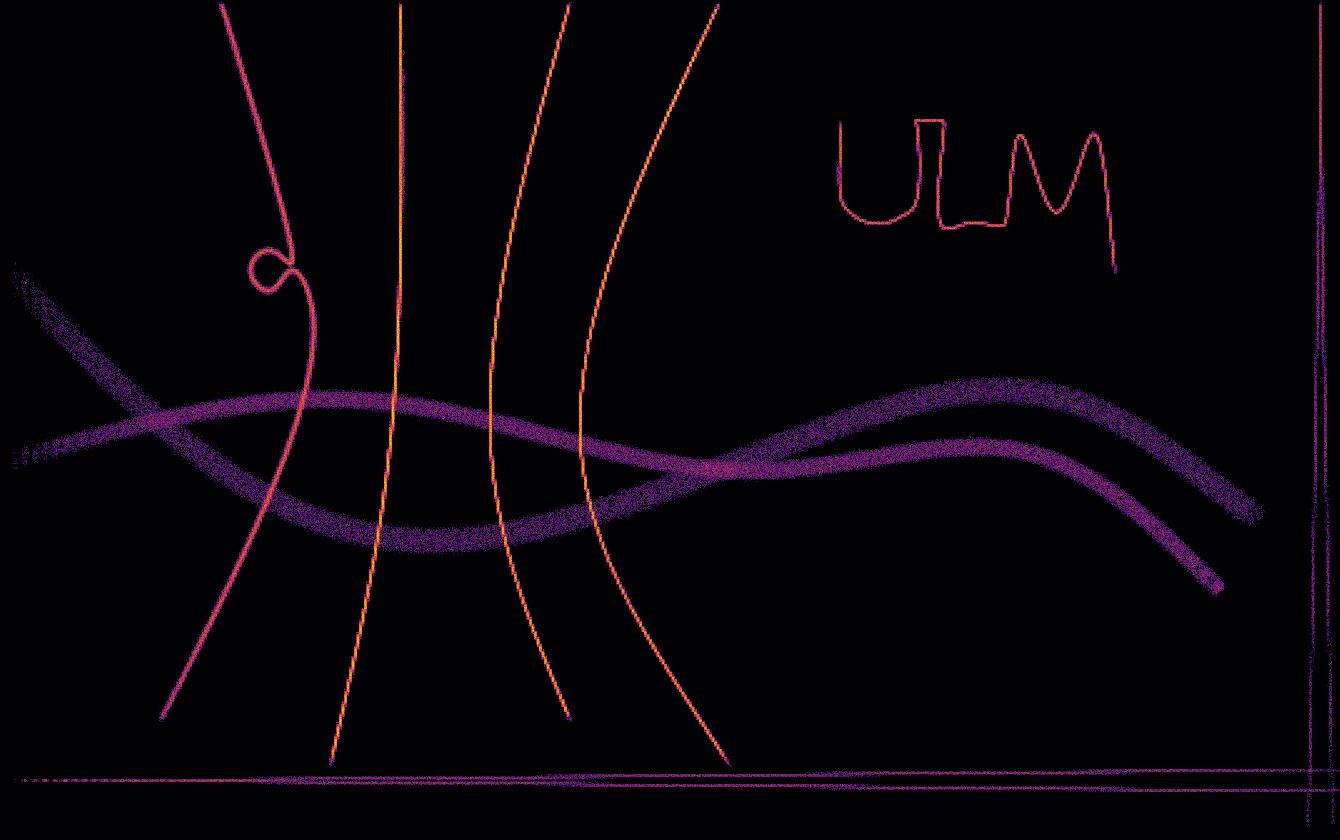}\par\vspace{1mm}
    \includegraphics[width=\textwidth,trim=310 45 980 750, clip]{./img/pala_sim_results/unet+iq_4.png}\par\vspace{1mm}
    \includegraphics[width=\textwidth,trim=820 560 200 100, clip]{./img/pala_sim_results/unet+iq_4.png}
    \subcaption{U-Net~\{128\}~\cite{van2020super}\label{fig:method4}}
    \vspace{+.1cm}
  \end{minipage}
  \hfill
  \begin{minipage}[b]{0.16\textwidth}
    \centering
    \includegraphics[width=\textwidth,trim=240 538 1030 238, clip]{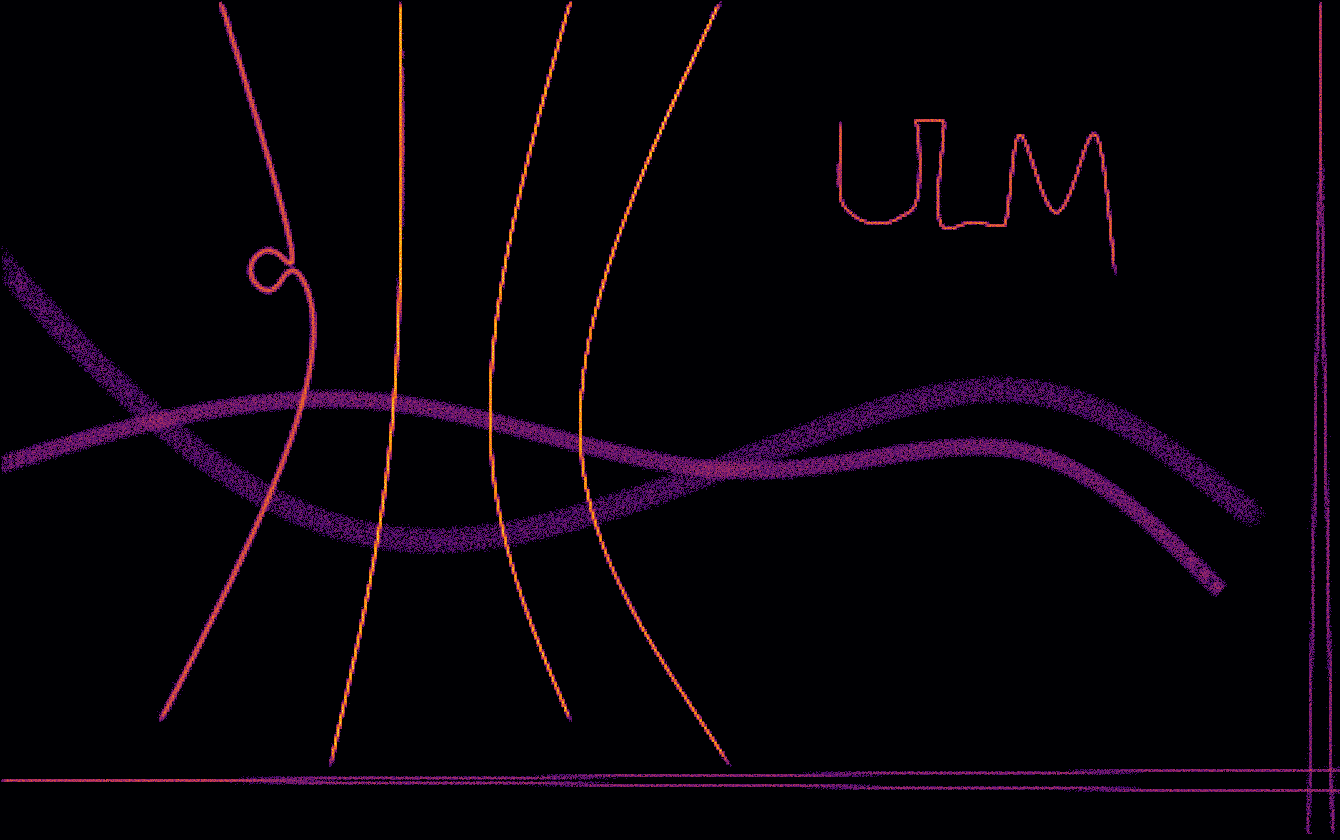}\par\vspace{1mm}
    \includegraphics[width=\textwidth,trim=310 45 980 750, clip]{./img/pala_sim_results/mspcn+iq_4.png}\par\vspace{1mm}
    \includegraphics[width=\textwidth,trim=820 560 200 100, clip]{./img/pala_sim_results/mspcn+iq_4.png}
    \subcaption{mSPCN~\{128\}~\cite{liu2020deep}\label{fig:method5}}
    \vspace{+.1cm}
  \end{minipage}
  \hfill
  \begin{minipage}[b]{0.16\textwidth}
    \centering
    \includegraphics[width=\textwidth,trim=240 538 1030 238, clip]{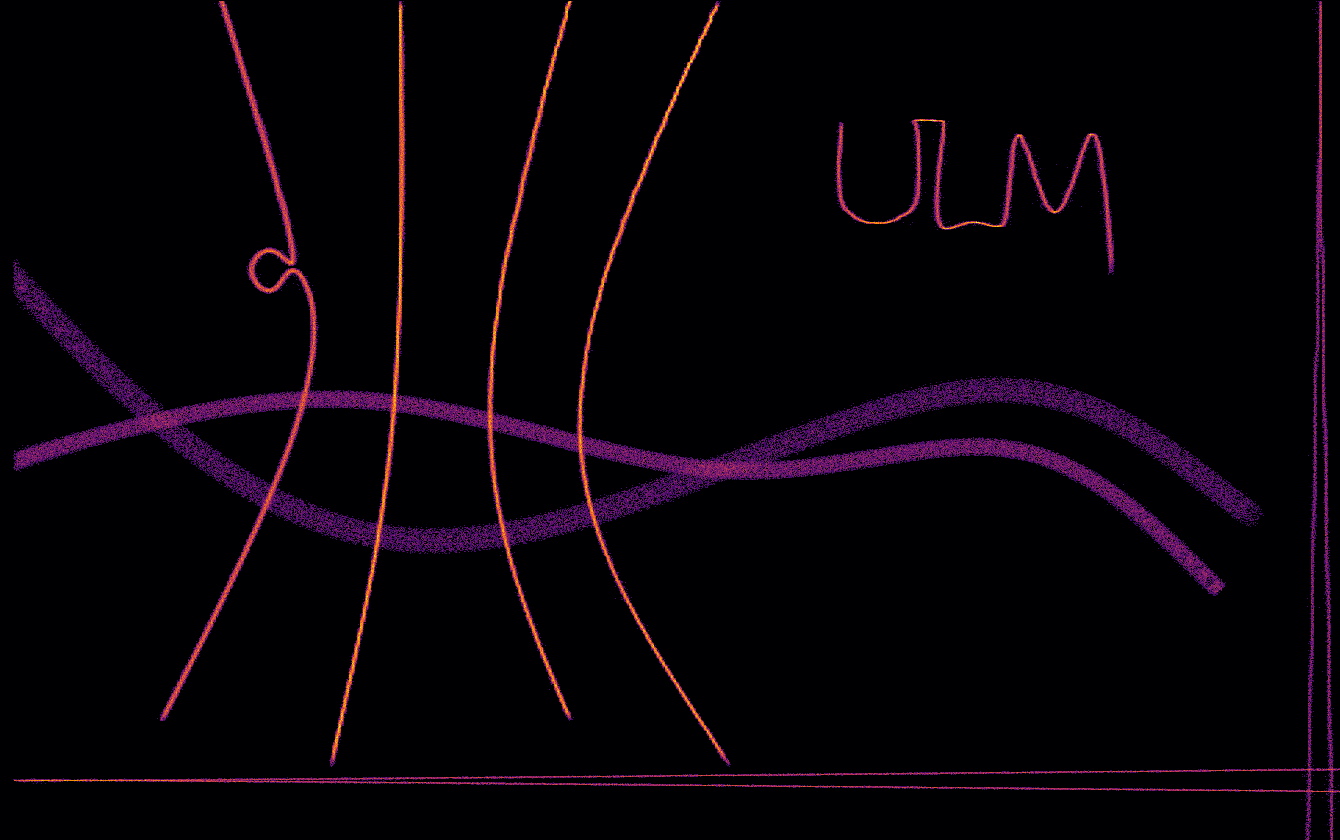}\par\vspace{1mm}
    \includegraphics[width=\textwidth,trim=310 45 980 750, clip]{./img/pala_sim_results/mspcn+rf_4.png}\par\vspace{1mm}
    \includegraphics[width=\textwidth,trim=820 560 200 100, clip]{./img/pala_sim_results/mspcn+rf_4.png}
    \subcaption{Ours~\{128\}\label{fig:method6}}
    \vspace{+.1cm}
  \end{minipage}
    \caption{\textbf{In silico ULM regions} from  Table~\ref{tab:benchmark} without temporal tracking. The methods in (\subref{fig:method2})-(\subref{fig:method3}) are deterministic approaches whereas (\subref{fig:method4})-(\subref{fig:method6}) are based on deep learning networks. While methods generally work on B-mode images, the methods in (\subref{fig:method3}) and (\subref{fig:method6}) received RF data as inputs.} 
    \label{fig:render_ulm_crop}
\end{figure*}
%
Image-based localization is a well-studied task in the computer vision domain. Researchers in the field of ULM recently adopted advances from DNNs such as the U-Net~\cite{van2020super} or mSPCN~\cite{liu2020deep} as a flavored super-resolution DNN. Super-resolution in DNNs is accomplished by a shuffle operation that learns to map feature channels to spatially upsampled images~\cite{liu2020deep}.

\textbf{Training:} For our network $f(\cdot)$, we employ the available RF frames from the PALA study~\cite{heiles2022pala} splitting the in silico set into 5000 training frames. We choose our RF-network architecture to be mSPCN~\cite{liu2020deep} with upsample factor $R=4$ for direct comparison with its B-mode counterpart. We train networks similar to~\cite{van2020super,liu2020deep}, however, bypass the preceding beamforming operation.
The loss function $\mathcal{L}(\cdot)$ for an RF input $\mathbf{X}\in\left[-1,1\right]^{2\times U\times V}$ and label mask $\mathbf{Y}\in\{0,1\}^{RU\times RV}$ is defined as follows:
\begin{align}
\mathcal{L}(\mathbf{X},\mathbf{Y})= \lVert f(\mathbf{X}) - \lambda_0(\mathbf{G}_{\sigma} \circledast \mathbf{Y})\rVert_2^2 + \lambda_1 \lVert f(\mathbf{X})\rVert_1
\end{align}
Here, $\circledast$ denotes the convolution operator, and $\lambda_0$ is a parameter that amplifies the labels. The second term of the loss function is an $L_1$ regularization term, scaled by $\lambda_1$, which prevents $f(\cdot)$ from predicting an excessive number of false positives. Other than existing studies, we provide all networks with complex numbers stacked as feature channels feeding more information as compared to magnitude-only input signals.
Our model undergoes training with an Adam optimizer, employing a batch size of 4, weight decay set at $1\mathrm{e}{-8}$, and an initial learning rate of $1\mathrm{e}{-3}$. The learning rate schedule is implemented using cosine annealing. For regularization purposes, the scaling factors are chosen as follows: $\lambda_0=(\max(\mathbf{G}_{\sigma}\circledast\mathbf{Y})/120)^{-1}$ and $\lambda_1=1\mathrm{e}{-2}$. The training process continues for a maximum of 40 epochs.
To enhance generalization and prevent overfitting, we incorporate input signal augmentation techniques. These include amplitude normalization, and the addition of clutter noise~\cite{heiles2022pala} with a clutter noise ratio of $50~\text{dB}$.

\textbf{Inference:}
Prior to network inference, spatio-temporal filtering is applied using Singular Value Decomposition (SVD) and a bandpass~\cite{heiles2022pala}. A frame predicted by $f(\mathbf{X})$ provides localization probabilities at each coordinate in an equidistant sampling grid. We extract point coordinates through thresholding of a predicted mask that underwent Non-Maximum Suppression (NMS) in advance. The ideal threshold largely depends on the learned network and dataset and is determined using \mbox{G-means} analysis of the ROC curve. NMS-based thresholding of localization probability maps yields maxima at upsampled integer coordinates, which are scaled to the original input resolution. It is important to note that these coordinates reflect localizations in RF space, which are required to be transferred to B-mode space for the finally rendered image.

\textbf{Forward Label Projection:}
%
To relieve the burden of the computational complexity imposed by Geometric ULM (G-ULM)~\cite{gulm:2023}, we project B-mode points to RF space and remap RF coordinates back to B-mode space using affine transformation algebra. Ground Truth (GT) labels are generally provided in \mbox{B-mode} coordinate space. However, learning localization directly from RF data requires to map these labels to RF coordinate space. We accomplish \mbox{B-mode} to RF data projection based on the physical Time-of-Flight (ToF) modelling. 
Given GT point labels $\mathbf{p}_i=\left[y_i,z_i,1\right]^\intercal$ with index $i$ in \mbox{B-mode} space, we project labels to a RF positions by 
\begin{align}
      %
      \mathbf{p}_{i,k}'= \Big(\lVert\mathbf{p}_i-\mathbf{v}_s\rVert_2+\lVert\mathbf{p}_i-\mathbf{x}_k\rVert_2-s\Big)\frac{f_s}{c_s}  
      \, , \quad \forall k \, ,
      \label{eq:project}
\end{align}
where $\mathbf{v}_s\in\mathbb{R}^3$ is the virtual source, $\mathbf{x}_k\in\mathbb{R}^3$ is a transducer position with index $k\in\{1,2,\dots,K\}$ and $\lVert\cdot\rVert_2$ is the Euclidean norm. Here, $s$ deducts the travel distance for the elapsed time between emission and capture start. The scalars $c_s$ and $f_s$ represent the speed of sound and sample rate, respectively.

Equation \eqref{eq:project} demonstrates that a single \mbox{B-mode} point $\mathbf{p}_i$ yields one label $\mathbf{p}'_{i,k}$ per channel $k$ in RF space. These points represent the wavefront that bounced back from a microbubble and would be merged to a single point distribution during DAS beamforming. To obtain GT labels, we isolate RF projections along the transducer dimension using
\begin{align}
    y_i^\star = \underset{k}{\operatorname{arg\,min}} \left\{y_{i,k}'\right\} \,,
    \quad \text{and} \quad 
    z_i^\star = \underset{k}{\operatorname{min}}\left\{z_{i,k}'\right\} \,,
\end{align}
which serve as RF training labels $\mathbf{p}_i^\star=\left[y_i^\star,z_i^\star,1\right]^\intercal$. 

It is crucial to avoid points $\mathbf{p}_i^\star$ being considered as GT labels if they are projected outside the transducer width. Similarly, these points are neglected as estimates for an inverse transformation mapping, which is introduced hereafter.

\textbf{Inverse Point Transformation:}
After inference and localization, we wish to remap localizations from RF space back to \mbox{\mbox{B-mode}} coordinates for visualization and comparison purposes. An analytical inverse of~\eqref{eq:project} turns out to be infeasible due to the Euclidean distance reduction. Instead, we model the reverse point mapping by an affine transformation defined as %
\begin{align}
    \begin{bmatrix}
    y_i \\
    z_i \\
    1 \\
    \end{bmatrix}
    &=
    \begin{bmatrix}
    a_{11} & a_{12} & a_{13} \\
    a_{21} & a_{22} & a_{23} \\
    0 & 0 & 1 \\
    \end{bmatrix}
    \begin{bmatrix}
    y_i^\star \\
    z_i^\star \\
    1 \\
    \end{bmatrix}
    \, ,
\end{align}
where $\mathbf{a}=\left[a_{11}, a_{12}, a_{13}, a_{21}, a_{22}, a_{23}\right]$ and $\mathbf{A}=g(\mathbf{a})$ with $g(\cdot):\mathbb{R}^6\mapsto\mathbb{R}^{3\times3}$. The coefficients $a_{11}, a_{12}, a_{21}, a_{22}$ take care of the affine scaling and shearing while $a_{13}, a_{23}$ are translation parameters. We employ the Levenberg-Marquardt method for an iterative least-squares optimization of $\mathbf{a}$ using 
\begin{align} \underset{\mathbf{a}}{\min} \, \,
\left\{\rVert\mathbf{A}\mathbf{p}_i^\star-\mathbf{p}_i\rVert_2^2\right\} \, ,
\end{align}
as the objective function. For the regression, we rely on synthetic random data points $\mathbf{p}_i$ in \mbox{B-mode} space with indices $i\in\{1,2,\dots,N\}$ while $N\gg6$. These synthetic points are projected to RF space using \eqref{eq:project} such that $\mathbf{A}$ is acquired once in advance and independent of training and inference. Note that coherent compounding requires to estimate a single transformation matrix $\mathbf{A}$ for each wave send direction. 
We fuse points over compounded waves via DBSCAN clustering with an eps of 0.5 wavelength units and a minimum cluster sample size of 1.

\section{Experiments}
For results assessment, we use established metrics from prior work~\cite{liu2020deep,heiles2022pala}. To gauge localization accuracy, we calculate the minimum Root Mean Squared Error (RMSE) between estimated and ground truth positions. RMSE values smaller than a quarter of the wavelength are treated as true positives contributing to the overall RMSE across frames~\cite{heiles2022pala}. Larger RMSE values of estimated positions are considered as false positives. Ground truth locations without an estimate within the wavelength threshold are marked as false negatives. To assess detection reliability, we use the Jaccard Index, which considers true positives, false positives and false negatives. Additionally, we report parameter count per model and inference time with a batch size of 1.

Table~\ref{tab:benchmark} presents results from the insilico PALA dataset~\cite{heiles2022pala} where networks have learned to upsample by factor $R=4$. 
%
The corresponding image regions are provided in Fig.~\ref{fig:render_ulm_crop} at 10 times higher resolution. For qualitative comparison with other methods, we introduce additive noise for ULM rendering to work against the coordinate quantization from NMS. This gives a more natural visual appearance than for example bicubic interpolation. 
The noise is uniform and within half the localization pixel size to not affect image quality. %
%
\begin{table}[!h]
	\centering
	\caption{Summary of localization results using 15000 frames of the PALA insilico data~\cite{heiles2022pala} and $R=4$. Metrics are reported as mean±std. where applicable. Units are given in brackets.}\label{tab:benchmark}
	\begin{tabular}{llcl}
		\toprule
		Method & 
        RMSE [$\lambda/10$] $\downarrow$ & Jaccard [$\%$] $\uparrow$ & Time [s] $\downarrow$ 
        \\
		\midrule
        Weighted Avg.~\cite{heiles2022pala} & 
        $1.287\pm0.162$ & 44.253 & 0.080 + $T_{\text{DAS}}$ 
        \\
		2-D Gauss Fit~\cite{song:2018} & 
        $1.240 \pm 0.162$ & 51.342 & 3.782 + $T_{\text{DAS}}$ 
        \\
		RS~\cite{heiles2022pala} & 
        $1.179 \pm 0.172$ & 50.330 & 0.099 + $T_{\text{DAS}}$ 
        \\ 
		  G-ULM~\cite{gulm:2023} & 
        $0.967\pm0.109$ & 78.618 & 3.747 
        \\
		\hline
		U-Net~\cite{van2020super} & 
        $0.950 \pm 0.084$ & 87.883
 & 0.017 + $T_{\text{DAS}}$ 
        \\
        mSPCN~\cite{liu2020deep} & 
        $0.978 \pm 0.085$ & 93.748 & 0.003 + $T_{\text{DAS}}$ 
        \\
		  \textbf{Ours} & 
        $0.858\pm0.137$ & 88.538 & 0.010 
        \\
		\bottomrule
	\end{tabular}
\end{table}

The accurate localization of our RF trained network is believed to be due to the following reasons: The wavefront distributions in RF data provide more spatial information allowing the network to make better predictions from the geometric shape. More precisely, the hyperbolic curvatures appearing in RF space help guiding the network weights to find the tip of an arriving wavefront. The spatial B-mode resolution in the PALA~\cite{heiles2022pala} frames is rendered to $143\times84$ pixels from the original $128\times256$ RF samples. This resampling involves downsampling the depth dimension while upsampling the lateral domain and is believed to alter localization performance. Besides, mSPCN has shown to learn real and imaginary numbers provided in the RF feature channels whereas learning the same from B-mode images did not significantly affect the localization metrics. %
Given the beamforming time $T_{\text{DAS}}$, 
the target computation time is smaller than the acquisition frame rate. The mSPCN model enables rapid computation of $15000$ frames within a reasonable time interval of 1-4 minutes on an Nvidia RTX 3090, which was used for training and inference.%
%
%
\begin{figure*}[!th]
    \centering
    \begin{minipage}[t]{\textwidth}
        \includegraphics[width=\linewidth,trim=18 38 15 6, clip]{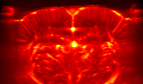}
        \subcaption{\label{fig:rat_iq}}
        \vspace{0.06cm}
    \end{minipage}
    \vfill
    \begin{minipage}[t]{\textwidth}
        \includegraphics[width=\linewidth,trim=180 380 70 60, clip]{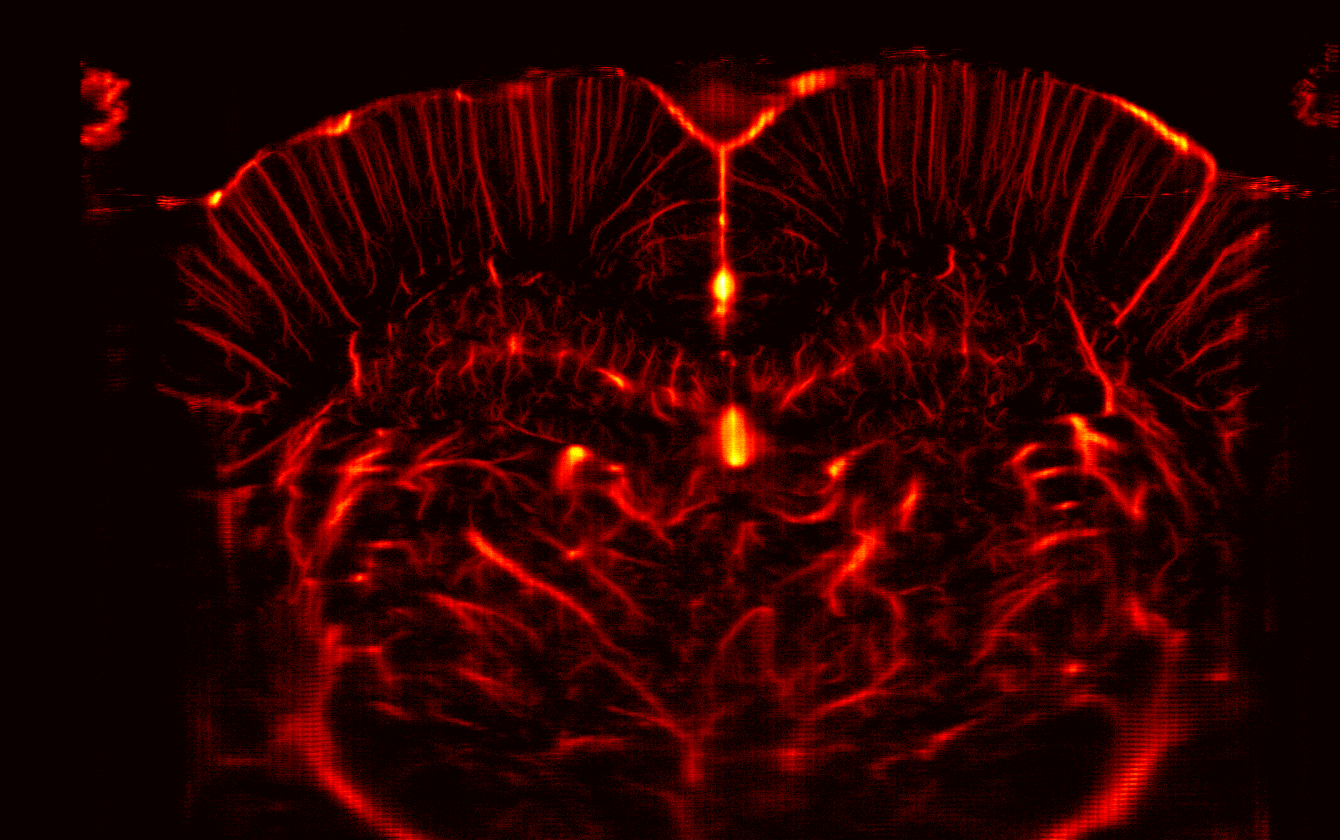}
        \subcaption{\label{fig:rat_rf}}
    \end{minipage}
    \caption{\textbf{In vivo results} showing the vascular structure of a rat brain. The image in (\subref{fig:rat_iq}) is reconstructed by averaging B-mode images after DAS beamforming whereas (\subref{fig:rat_rf}) shows accumulated localizations from our RF-trained network with $R=10$. The methods employ 128 transducer channel data from $120\times800$ frames without temporal tracking.}
\end{figure*}
\section{Conclusion}
%
This paper demonstrates the feasibility of localizing microbubbles without the need for delay and sum beamforming. This achievement is accomplished through the utilization of a super-resolution deep neural network in tandem with custom forward and backward transformations, facilitating the mapping of points between RF and B-mode coordinate spaces. An existing network architecture has successfully acquired the ability to accurately pinpoint the tip of an incoming wavefront. 
Our study not only reveals that the omission of beamforming reduces computational complexity but also enhances state-of-the-art localization accuracy and detection reliability. These improvements are substantiated by numerical quantification from an available dataset. We are confident that these findings will significantly contribute to the advancement of future ULM pipelines, ultimately paving the way for the clinical adoption of this promising technology.

\bibliographystyle{IEEEtran}
\bibliography{main}

\end{document}